\documentclass[
 reprint,
 superscriptaddress,
 amsmath,
 amssymb,
 aps,
 floatfix,
 longbibliography,  %
]{revtex4-2}

\usepackage{bm} %
\usepackage{booktabs} %
\usepackage{graphicx}  %
\usepackage[colorlinks,citecolor=blue,urlcolor=blue]{hyperref} %
\usepackage{siunitx} %
\usepackage{xcolor} %
\usepackage{todonotes} %
\setuptodonotes{inline} %

\usepackage[framemethod=TikZ]{mdframed}  %
\newcounter{boxcnt}
\newenvironment{boxenv}[1]
    {\begin{figure*}[tb]
    \refstepcounter{boxcnt}
    
    \begin{mdframed}[roundcorner=10pt,backgroundcolor=blue!10]
    \textbf{\centerline{Box \arabic{boxcnt}: #1}}
    \smallskip

    }
    {\end{mdframed}\end{figure*}
    }

\newcommand{\figref}[1]{\mbox{Fig.\hspace{0.25em}\ref{#1}}}

\newcommand{\secref}[1]{\mbox{section\hspace{0.25em}\ref{#1}}}

\begin{document}

\title{Roadmap for Condensates in Cell Biology}

\author{Dilimulati Aierken}
\affiliation{Department of Chemical and Biological Engineering, Princeton University, NJ 08544, USA}
\affiliation{Omenn-Darling Bioengineering Institute, Princeton University, NJ 08544, USA}

\author{Sebastian Aland}
\affiliation{Institute of Numerical Mathematics and Optimization, TU Bergakademie Freiberg, 09599 Freiberg, Germany}
\affiliation{Faculty of Informatics/Mathematics, HTW Dresden, 01069 Dresden, Germany}

\author{Stefano Bo}
\affiliation{Department of Physics, King's College London, London WC2R 2LS, United Kingdom}

\author{Steven Boeynaems}
\affiliation{Department of Molecular and Human Genetics, Baylor College of Medicine, Houston, TX 77030, USA}
\affiliation{Jan and Dan Duncan Neurological Research Institute, Texas Children’s Hospital, Houston, TX 77030, USA}
\affiliation{Therapeutic Innovation Center (THINC), Baylor College of Medicine, Houston, TX 77030, USA}
\affiliation{Center for Alzheimer's and Neurodegenerative Diseases (CAND), Baylor College of Medicine, Houston, TX 77030, USA}
\affiliation{Dan L Duncan Comprehensive Cancer Center (DLDCCC), Baylor College of Medicine, Houston, TX 77030, USA}

\author{Danfeng Cai}
\affiliation{Department of Biochemistry and Molecular Biology, Johns Hopkins Bloomberg School of Public Health, Baltimore, Maryland}
\affiliation{Department of Biophysics and Biophysical Chemistry, Johns Hopkins School of Medicine, Baltimore, Maryland}
\affiliation{Department of Oncology, Johns Hopkins School of Medicine, Baltimore, Maryland}

\author{Serena Carra}
\affiliation{Department of Biomedical, Metabolic and Neural Sciences, University of Modena and Reggio Emilia, Modena, 41125, Italy}

\author{Lindsay B. Case}
\affiliation{Department of Biology, Massachusetts Institute of Technology, Cambridge, MA, 02139, USA}

\author{Hue Sun Chan}
\affiliation{Department of Biochemistry, University of Toronto, Toronto, Ontario M5S 1A8, Canada}

\author{Jorge R. Espinosa}
\affiliation{Department of Physical Chemistry, University Complutense of Madrid, Madrid, Spain, 28040}
\affiliation{Instituto Pluridisciplinar, Universidad Complutense de Madrid, P.\textordmasculine{} de Juan XXIII, 1, Moncloa - Aravaca, 28040 Madrid, Spain}
\affiliation{Yusuf Hamied Department of Chemistry, University of Cambridge, Lensfield Road, Cambridge CB2 1EW, UK}

\author{Trevor K. GrandPre}
\affiliation{Department of Physics and Center for Biomolecular Condensates, Washington University in St. Louis, St. Louis, MO 63130, USA}
\affiliation{National Institute for Theory and Mathematics in Biology, Northwestern University and The University of Chicago, Chicago, IL, USA}

\author{Alexander Y. Grosberg}
\affiliation{Department of Physics and Center for Soft Matter Research, New York University, 726 Broadway, New York, NY 10003, USA}

\author{Ivar S. Haugerud}
\affiliation{Faculty of Mathematics, Natural Sciences, and  Engineering: Institute of Physics, University of Augsburg, Universitätsstrasse 1, 86159 Augsburg, Germany}

\author{William M. Jacobs}
\affiliation{Department of Chemistry, Princeton University, Princeton, NJ 08544, USA}

\author{Jerelle A. Joseph}
\thanks{Contact author: jerellejoseph@princeton.edu}
\affiliation{Department of Chemical and Biological Engineering, Princeton University, NJ 08544, USA}
\affiliation{Omenn-Darling Bioengineering Institute, Princeton University, NJ 08544, USA}

\author{Frank Jülicher}
\thanks{Contact author: julicher@pks.mpg.de}
\affiliation{Max Planck Institute for the Physics of Complex Systems, Nöthnitzer Str. 38, 01187 Dresden, Germany}

\author{Kurt Kremer}
\affiliation{Max Planck Institute for Polymer research, Ackermannweg 10, 55128 Mainz, Germany}

\author{Guido Kusters}
\affiliation{Max Planck Institute for Dynamics and Self-Organization, Am Faßberg 17, 37077 Göttingen, Germany}

\author{Liedewij Laan}
\affiliation{Department of Bionanoscience, Kavli Institute of Nanoscience Delft, Delft University of Technology, Delft, The Netherlands}

\author{Keren Lasker}
\affiliation{Department of Integrative Structural and Computational Biology, The Scripps Research Institute, La Jolla, CA, 92037, USA}

\author{Kathrin S. Laxhuber}
\affiliation{Max Planck Institute for the Physics of Complex Systems, Nöthnitzer Str. 38, 01187 Dresden, Germany}

\author{Hyun O. Lee}
\thanks{Contact author: hyunokate.lee@utoronto.ca}
\affiliation{Department of Biochemistry, University of Toronto, Toronto, Ontario, M5G1M1, Canada}

\author{Kathy F. Liu}
\affiliation{Department of Biochemistry and Biophysics, Perelman School of Medicine, University of Pennsylvania, Philadelphia, PA 19104, USA}
\affiliation{Graduate Group in Biochemistry and Molecular Biophysics, Perelman School of Medicine, University of Pennsylvania, Philadelphia, PA 19104, USA}
\affiliation{Abramson Family Cancer Research Institute, Perelman School of Medicine, University of Pennsylvania, Philadelphia, PA 19104, USA}
\affiliation{Penn Institute for RNA Innovation, University of Pennsylvania, Philadelphia, PA 19104, USA}
\affiliation{Penn Center for Genome Integrity, University of Pennsylvania, Philadelphia, PA 19104, USA}

\author{Dimple Notani}
\affiliation{National Centre for Biological Sciences-TIFR, GKVK Campus, Bellary Road, Bangalore, India. 560065}

\author{Yicheng Qiang}
\affiliation{Max Planck Institute for Dynamics and Self-Organization, Am Faßberg 17, 37077 Göttingen, Germany}

\author{Paul Robustelli}
\affiliation{Department of Chemistry, Dartmouth College, Hanover, NH, 03755, USA}

\author{Leonor Saiz}
\affiliation{Department of Biomedical Engineering, University of California, Davis, CA 95616, USA}

\author{Omar A. Saleh}
\affiliation{Materials and Physics Departments, University of California, Santa Barbara, CA 93110, USA}

\author{Helmut Schiessel}
\affiliation{Cluster of Excellence Physics of Life, TU Dresden, 01062 Dresden, Germany}

\author{Jeremy Schmit}
\affiliation{Department of Physics, Kansas State University, Manhattan, KS 66506, USA}

\author{Meng Shen}
\affiliation{Department of Physics, California State University, Fullerton, CA 92831, USA}

\author{Krishna Shrinivas}
\affiliation{Department of Chemical and Biological Engineering, Northwestern University, Evanston IL 60208, USA}
\affiliation{Center for Synthetic Biology, Northwestern University, Evanston IL 60208, USA}

\author{Antonia Statt}
\affiliation{Department of Materials Science and Engineering, University of Illinois Urbana-Champaign, Urbana, Illinois 61801, USA}

\author{Andres R. Tejedor}
\affiliation{Department of Physical Chemistry, University Complutense of Madrid, Madrid, Spain, 28040}
\affiliation{Instituto Pluridisciplinar, Universidad Complutense de Madrid, P.\textordmasculine{} de Juan XXIII, 1, Moncloa - Aravaca, 28040 Madrid, Spain}
\affiliation{Yusuf Hamied Department of Chemistry, University of Cambridge, Lensfield Road, Cambridge CB2 1EW, UK}

\author{Tatjana Trcek}
\affiliation{Department of Biology; Johns Hopkins University; Baltimore, MD 21218, USA}

\author{Christoph A. Weber}
\affiliation{Faculty of Mathematics, Natural Sciences, and  Engineering: Institute of Physics, University of Augsburg, Universitätsstrasse 1, 86159 Augsburg, Germany}

\author{Stephanie C. Weber}
\thanks{Contact author: steph.weber@mcgill.ca}
\affiliation{Department of Biology, McGill University, Montreal, H3A 1B1, Canada}
\affiliation{Department of Physics, McGill University, Montreal, H3A 2T8, Canada}

\author{Ned S. Wingreen}
\thanks{Contact author: wingreen@princeton.edu}
\affiliation{Department of Molecular Biology, Princeton University, NJ 08544, USA}
\affiliation{Lewis-Sigler Institute for Integrative Genomics, Princeton University, NJ 08544, USA}

\author{Huaiying Zhang}
\affiliation{Department of Biological Sciences, Carnegie Mellon University, Pittsburgh, PA, 15213, USA}

\author{Yaojun Zhang}
\affiliation{Department of Physics \& Astronomy, Johns Hopkins University, Baltimore, MD 21218, USA}
\affiliation{Department of Biophysics, Johns Hopkins University, Baltimore, MD 21218, USA}

\author{Huan Xiang Zhou}
\affiliation{Department of Chemistry and Department of Physics, University of Illinois Chicago, Chicago, IL 60607, US}

\author{David Zwicker}
\thanks{Contact author: david.zwicker@ds.mpg.de}
\affiliation{Max Planck Institute for Dynamics and Self-Organization, Am Faßberg 17, 37077 Göttingen, Germany}

\begin{abstract}
    Biomolecular condensates govern essential cellular processes yet elude description by traditional equilibrium models. This roadmap, distilled from structured discussions at a workshop and reflecting the consensus of its participants, clarifies key concepts for researchers, funding bodies, and journals. After unifying terminology that often separates disciplines, we outline the core physics of condensate formation, review their biological roles, and identify outstanding challenges in nonequilibrium theory, multiscale simulation, and quantitative in-cell measurements. We close with a forward-looking outlook to guide coordinated efforts toward predictive, experimentally anchored understanding and control of biomolecular condensates.
\end{abstract}

\maketitle

\section{Introduction}\label{sec:introduction}

\begin{figure}[b]
    \centering
    \includegraphics[width=\linewidth]{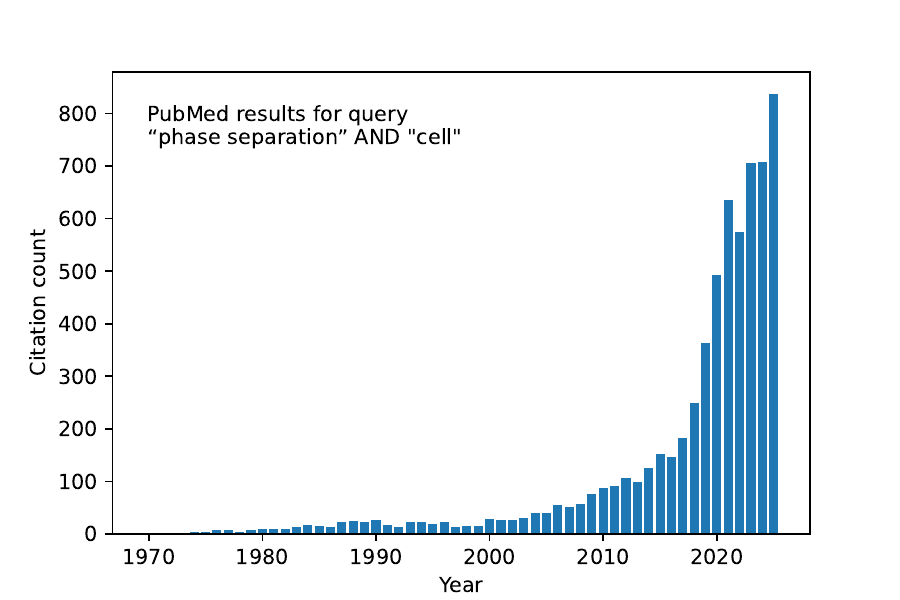}
    \caption{
    The strong rise in the number of publications in recent years (obtained from searching PubMed for “phase separation” and “cell”) indicates a growing interest in using the concept of phase separation in cell biology.
    }
    \label{fig:citations}
\end{figure}

Biomolecular condensates have emerged as a key organizing principle in cell biology, transforming our understanding of cellular architecture (\figref{fig:citations}). Recent studies have highlighted that condensates are involved in a wide range of essential processes, including gene regulation, signal transduction, metabolism, macromolecular processing, signaling, and stress response. Beyond fundamental biology, condensates are linked to human diseases such as neurodegeneration and cancer, as well as to climate-impacting processes such as carbon fixation, underscoring their widespread significance. To fully understand how cells operate, we need to understand condensates, which, in turn, will enable us to target or engineer them for therapeutic and environmental benefit. The concepts of phase separation and condensation offer a powerful framework for understanding these non-membrane-bound organelles, providing insight into how they organize biomolecules in space and time. Understanding biological condensates requires integrating concepts from biology, physics, and chemistry, and collaboration across disciplines is essential to advance the field.

\begin{figure*}
    \centering
    \includegraphics[width=\linewidth]{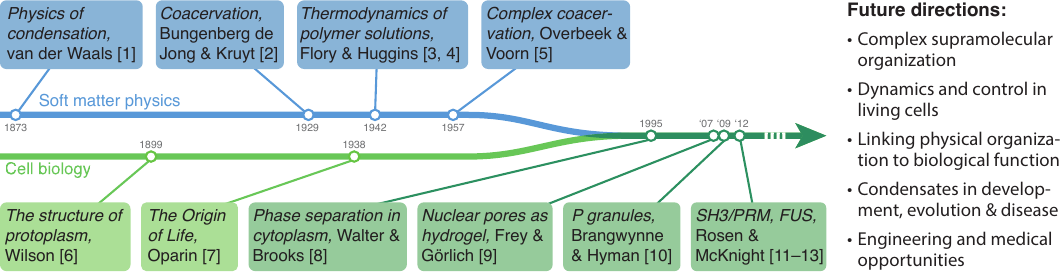}
    \caption{
    \textbf{Historical development of condensates in cell biology.}
    The timeline highlights selected landmark papers that influenced biomolecular condensates research substantially. Future directions are discussed in detail in \secref{sec:outlook}.
    }
    \nocite{van1873over, Jong1929, Flory1942, Huggins1941, Overbeek1957} %
    \nocite{Wilson1899, Oparin1938, Walter1995, Frey2007, Brangwynne2009, li_phase_2012, kato2012, han_cell-free_2012} %
    \label{fig:timeline}
\end{figure*}

The concept of condensation has deep roots in physics, dating back over a century (\figref{fig:timeline}). Initially developed to explain phase separation in thermodynamic equilibrium, these concepts have been continuously refined to describe more complex situations in soft matter physics. Similar principles were invoked in biology long before any molecular details were understood: early observations of cellular space hinted at numerous liquid-like self-assemblies without membranes. Advances in molecular biology and imaging have since revealed the complex composition and dynamics of these structures, bridging phenomenology with underlying mechanisms in increasing detail (\figref{fig:timeline}). Today, the combination of soft matter physics and cell biology forms a truly interdisciplinary field that offers an opportunity for discovery for all. There are a growing number of success stories in this regard; examples include nucleoli that generate protein-synthesizing ribosomes, germ granules that specify developing germ cells, the asymmetric organization of cell-division components in bacteria, and recombination nodules (Box~\ref{box}). 

\begin{figure*}
    \centering
    \includegraphics[width=\linewidth]{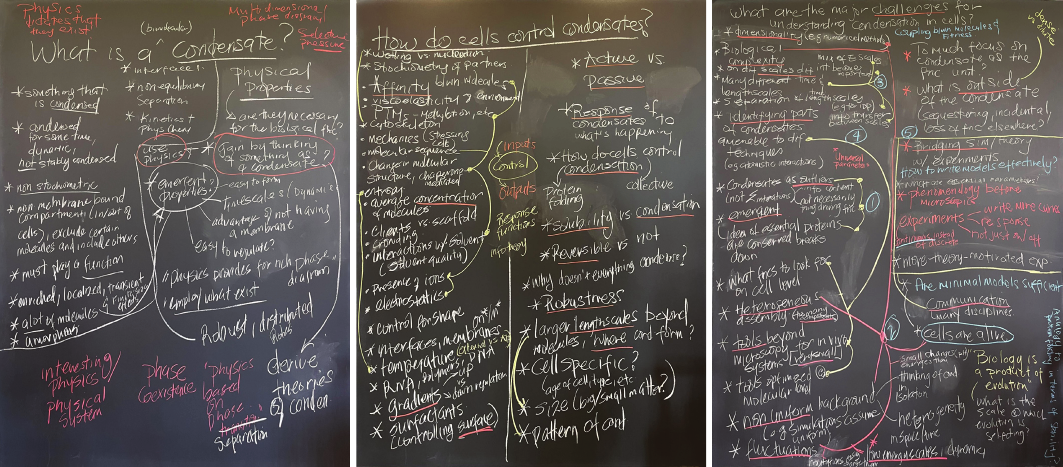}
    \caption{
    Pictures of blackboards summarizing weekly discussion sessions at the KITP Workshop “Physical Principles Shaping Biomolecular Condensates” in 2025.
    The three shown examples are related to sections \ref{sec:terminology}, \ref{sec:physics}, and \ref{sec:challenges}.
    }
    \label{fig:boards}
\end{figure*}

This interdisciplinarity offers both opportunities for discovery and challenges. Scientists from biology, chemistry, physics, and mathematics often approach condensates with different languages and conceptual frameworks, making communication and efficient progress challenging. To address this, researchers convened at the Kavli Institute for Theoretical Physics (KITP) for seven weeks in the summer of 2025. Many of the \href{https://online.kitp.ucsb.edu/online/biomol25/}{scientific talks are available online}. In addition to these seminars, every Friday, structured two-hour discussion sessions brought all participants together to share perspectives and address key questions. Ideas were summarized live on the blackboard (\figref{fig:boards}) and recorded in detailed notes, which form the foundation of this article. 

This roadmap aims to clarify concepts for scientists, funding agencies, and journals working in the area of biomolecular condensate research, reflecting the consensus view of the participants of the KITP workshop. We begin by addressing terminology, which often creates barriers between disciplines (\secref{sec:terminology}). Next, we examine the physics behind biological condensates (\secref{sec:physics}), summarize their biological implications (\secref{sec:biology}), and conclude with key challenges (\secref{sec:challenges}) and a future outlook (\secref{sec:outlook}).

\section{What are condensates?}
\label{sec:terminology}
The term biomolecular condensate (herein called condensate) is controversial \cite{musacchio2022a}. Some might consider all cellular structures to be condensates since all structures comprise condensed matter, i.e., are made from materials ranging from liquids to solids. Others might claim that nothing is a condensate since they define condensate as emerging through a phase transition, which would require infinitely large systems to unambiguously identify them using tools from statistical mechanics. Clearly, neither of these viewpoints is productive for describing and understanding cell biology. Instead, people have developed a wealth of terms, like aggregates, blobs, bodies, clusters, foci, granules, hubs, macromolecular assemblies, microdomains, nodules, puncta, ribosome exclusion zones, speckles, spots, etc., to describe what might be summarized as biomolecular condensates. Such a zoo of terminology with subtly distinct meanings hinders coherent communication and suggests that clarifying the concept of condensates might be helpful.

What properties do condensates have in common? Although opinions differ, researchers tend to agree that condensates are compartments with a very different composition from their surroundings. The surroundings are separated by a well-defined interface that is not established by a lipid membrane; condensates are non-membrane-bound compartments that can exclude membrane-bound organelles such as the nucleus or mitochondria (sometimes they associate with one another). People also tend to agree that condensates have dynamic, non-stoichiometric composition, so individual complexes, like the ribosome, do not count (clusters of complexes may). Moreover, condensates are typically described as having (at least in part) an amorphous internal structure, in which the relative positions of molecules are not guaranteed, to exclude ordered structures such as amyloid-like aggregates and microtubules. Finally, many people stress that condensates exhibit emergent properties, i.e., that the properties of individual molecules differ from those of the collective. However, it is unclear at what size (or molecule count) such properties emerge (e.g., water exhibits liquid properties for as few as a dozen molecules \cite{rognoni_how_2021}). Consequently, there is no consensus on the size restrictions of condensates at this time.

There are also more controversial properties that are often discussed in the context of condensates: Some papers equate condensates with liquid-liquid phase separation (LLPS), but it is often neither clear whether the involved objects have liquid-like properties (and solid-like properties have been observed \cite{jawerth_protein_2020}), nor whether they actually form by phase separation. In any case, phase separation denotes the process, whereas the outcome is better described as phase coexistence. However, phase coexistence is unambiguously defined only in thermodynamically large equilibrium systems, so its validity in small, living cells out of equilibrium is unclear. It is also sometimes said that condensates are transient structures, although long-lived examples, like the Balbiani body, exist. Finally, condensates are often associated with biological function, but some condensates may form without a biological function \cite{putnam2023a}. In any case, proving biological function can be hard. We argue that it is still useful to discuss condensates even when their biological function is yet unclear, for example, to analyze structures observed in cells. Taken together, these controversies demonstrate that defining biomolecular condensates via their properties is virtually impossible.

We believe that it is infeasible to define the term "biomolecular condensate" unambiguously. In the end, biological structures exist along a continuum of sizes, material properties, and dynamics. Any sharp boundaries necessary to define a term are arbitrary and thus semantic. To make headway, we here suggest that one could instead think of condensation as a general concept and ask whether using it to understand a particular biological structure is useful: Can condensation explain observations (quantitatively)? Can it make testable predictions? In other words, we propose that the initial qualitative question (What are condensates?) is not productive since it easily leads to semantic discussions. Instead, we suggest reframing this to answer quantitative questions (e.g., how well does condensation explain a phenomenon?), which acknowledges the blurred nature of biology.

The concept of phase coexistence has certainly been useful to describe biological phenomena (Box~\ref{box}). Some of the examples remained mysterious for decades, and shifting the perspective to phase coexistence allowed describing them intuitively. In cases where such a description is successful, we might as well call the observed structure a condensate. In contrast, some structures currently called condensates might turn out to be better described by other concepts, in which case we suggest not calling them condensates anymore. In any case, phase coexistence is a generic property of complex mixtures, so this process likely governs some cellular structures. In other words, “condensation exists, so cells need to deal with it” (Alexander "Shura" Grosberg).

\section{Physics of condensates}
\label{sec:physics}

\begin{figure}
    \centering
    \includegraphics[width=\linewidth]{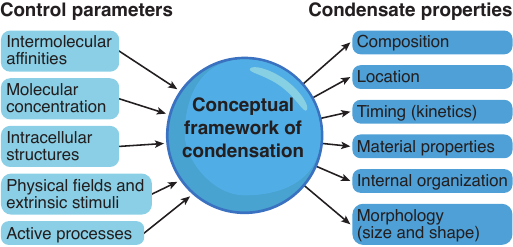}
    \caption{
    \textbf{The physical framework of condensation describes how various control parameters affect condensate properties.} Cells can use the control parameters as knobs to tune condensate properties. Alternatively, these properties serve as proxies for environmental parameters to perform sensing tasks.
    }
    \label{fig:concept}
\end{figure}

Recasting the question from “What is a condensate?” to the quantitative “Can the principles of condensation provide new tools to understand cellular phenomena?” focuses attention on control parameters and measurable responses. Practically, we ask which cellular “knobs” control condensation and how they shape the resulting responses. Treating biomolecular condensation as a continuous input--output mapping (from control parameters to responses) enables physically grounded, interpretable models that connect scattered observations and yield testable predictions (\figref{fig:concept}). In what follows, we first outline the control parameters that govern condensation in cells and then describe the condensate responses (i.e., measurable properties) that emerge.

\begin{boxenv}{Examples of success stories}
\label{box}

While many biomolecular condensates have been described \cite{kuznetsova_cd-code_2025}, the biological consequences of their formation remain unclear in most cases. Here, we highlight a few examples for which the concept of condensation has been especially useful for explaining biological phenomena.

\begin{itemize}
\item Among the first condensates described as such are \textbf{P granules}, which contribute to germ cell specification in the nematode \textit{Caenorhabditis elegans}. P granules are initially distributed throughout the one-cell embryo, but are ultimately inherited by the posterior daughter cell at division. This asymmetric segregation, first described in 1982 \cite{strome_immunofluorescence_1982}, cannot be explained by cytoplasmic flows or localized degradation. Instead, \citet{Brangwynne2009} proposed an alternative mechanism: before cell division, P granules in the anterior half of the embryo shrink and dissolve, while those in the posterior half condense and grow. This novel description explained not only the asymmetric localization of P granules in the embryo, but also their fusion, dripping, and wetting behavior in the adult gonad \cite{Brangwynne2009}. 

\item The bacterium \textit{Caulobacter crescentus} divides asymmetrically, despite a uniform distribution of the master regulator CtrA \cite{chen_spatial_2011}. Nevertheless, a gradient in CtrA activity is observed from the new pole to the old pole that is critical for cell fate determination. \citet{lasker2020} proposed that this activity gradient is established through selective sequestration of CtrA and the phosphosignalling proteins CckA and ChpT in the \textbf{Polar Organizing Protein Z (PopZ) microdomain}. Rational mutations that either fluidize or harden PopZ microdomains confer severe growth defects \cite{lasker2022}. Remarkably, combining these opposing mutations not only restores condensate fluidity but also rescues cell growth, demonstrating a direct relationship between condensate properties and organismal fitness. 

\item In sexually reproducing organisms, genetic diversity is generated through DNA exchange between homologous chromosomes during meiosis \cite{girard_regulation_2023}. Numerous double-strand breaks (DSBs) are formed along each chromosome, but only a few of these sites become crossovers. Indeed, putative crossover sites “interfere” with one another such that mature crossovers are distantly spaced. First observed in \textit{Drosophila} in 1913 \cite{sturtevant_linear_1913}, the mechanism underlying crossover interference remained a mystery for more than a century. Recent studies in the plant \textit{Arabidopsis thaliana} \cite{morgan_diffusion-mediated_2021} and the animal \textit{C. elegans} \cite{zhang_crossover_2025} now suggest that the number and positioning of crossovers are determined by coarsening of \textbf{recombination nodules}. Recombination nodules are condensates that form at DSBs. Nodule components, such as the RING finger protein HEI10/ZHP-3, diffuse laterally along the synaptonemal complex that holds homologous chromosomes together, redistributing from many small nodules to a few large nodules, which then promote crossover formation. Since nearby recombination nodules compete for material, the remaining large nodules tend to be distantly spaced.
\end{itemize}

Interestingly, although modulation of chemical reaction rates is the most commonly speculated function of condensates, these examples instead involve the precise localization of macromolecules on larger length scales \cite{lyon2020}. Indeed, biological consequences can occur on multiple length and time scales (i.e., molecular to evolutionary) and investigating condensates at the systems level may be necessary to identify emergent consequences.
\end{boxenv}

\subsection{Control Parameters (“Knobs”)}

We identified five broad classes of physical control parameters or knobs that affect condensates and could be used to influence their behavior.

\begin{enumerate}
\item\textbf{Intermolecular affinities.} These are encoded by the sequences of the biomolecules (proteins, RNA, DNA, etc.) that form condensates. For a given set of conditions, associative interactions among biomolecules can drive condensation by providing favorable binding free energy that compensates for the entropic losses associated with demixing. Chemical modifications (the addition of covalent tags) can reshape this interaction landscape; for example, post-translational modifications can drastically change the phase behavior of proteins and their ability to interact with RNA.  The sequence of molecules also shapes their structure, which in turn affects intermolecular affinities. For example, much of the studies has focused on interactions among intrinsically disordered protein sequences, which lack stable tertiary structure and can engage in multivalent, promiscuous associations that facilitate condensation. However, specific interactions are equally important. Repeating RNA motifs and folded protein domains (e.g., RNA-recognition motifs) can form high-affinity contacts that promote protein–RNA condensates; in some assemblies, structured-domain multimerization is the primary driver (e.g., SUMO-SIM \cite{hess_structured_2025}). Thus, it is crucial to consider the full spectrum of interaction types---from low-affinity, multivalent contacts to high-specificity binding---that give rise to condensation in cells.

\item\textbf{Molecular concentration.} Another important control parameter of condensation is molecular concentration. This encompasses the cellular concentrations of macromolecules, as well as small molecules and ions. Together, these components determine the effective solvent quality (e.g., pH, ionic strength) and thereby set the accessible phase space---i.e., the number and types of coexisting phases---of the “molecular mixture”. Additionally, molecular concentration directly determines the relative amounts of given macromolecules, thereby influencing their ability to form condensates. For example, at low RNA concentrations, certain proteins undergo condensation, whereas at sufficiently high RNA concentrations, protein condensation can be suppressed. Ion concentration strongly modulates electrostatic screening, thereby altering molecular affinities, especially for highly charged molecules such as RNA. As the molecular composition of the cellular milieu changes due to active processes (e.g., transcription, translation, degradation, and transport; discussed below), the accessible phase space changes, and with it the capacity of the multicomponent molecular mixture to condense into distinct phases.

\item\textbf{Cellular structures.} Cellular structures, including membranes, the cytoskeleton, membrane-bound organelles, and other surfaces, also dictate condensation inside cells. These structures can provide sites for heterogeneous nucleation, promote wetting or prewetting, impose curvature, and exert mechanical stresses, and thereby modulate condensate dynamics and shape. For example, attractive interactions with a surface can create prewetting layers that lower nucleation barriers and control droplet size and location. Conversely, surfaces (or surface-active molecules associated with them) can destabilize condensates by acting as surfactants that alter interfacial properties (e.g., reducing interfacial tension, changing contact angles) or by sequestering key components. Cytoskeletal networks can also position condensates, bias coalescence, and influence material exchange through active transport and forces.

\item\textbf{Physical fields (including extrinsic stimuli).} The final class of control parameters comprises physical fields---temperature, electric fields/potentials, hydrostatic pressure, pH, salt, cosolvents, and osmotic or mechanical stress---which generally act nonspecifically to shift phase boundaries and modulate condensation. These can be viewed as extrinsic cues that influence the phase behavior of molecules. 
Temperature, e.g., is a global control parameter: modest changes can move systems across the binodal. Within single cells, sustained spatial temperature gradients are thought to be minimal, but local energy dissipation and composition changes (e.g., ATP levels, ion fluxes, chaperone activity) can produce temperature-like effects. Electric fields (externally applied or arising from membrane potentials) redistribute ions and reweight electrostatic interactions, altering nucleation, wetting, and condensate dynamics. By contrast, gravity is negligible at most cellular length scales (though sedimentation/centrifugation can matter in large cellular compartments or in larger volume buffer solutions). Importantly, temperature also couples to sequence-dependent changes in molecular structure and solvation, creating a rich interplay between molecular sequence, interaction strengths, physical stimuli, and condensation propensities.

\item\textbf{Active processes.} Cells are living entities that metabolize material to avoid reaching thermodynamic equilibrium. Such active processes also affect condensates, e.g., by motor-driven active fluxes or active chemical conversions. The latter example allows modifying molecules into an unfavorable state, e.g., turning a molecule inside a condensate into a form that effectively repels the other condensate material, thus biasing this molecule to leave the condensate. Alternatively, molecules dissolved in the cytosol could be actively converted into a form that favors condensation, thereby initiating the formation of a condensate at a controlled location and at a specified time.
\end{enumerate}

These knobs rarely act in isolation; they are cross-coupled, producing rich biomolecular phase behavior. Moreover, some parameters are actively regulated by cells (e.g., post-translational modifications, local concentrations), others vary passively with physiology (e.g., macromolecular crowding, ion balance), and some are largely extrinsic (e.g., temperature, pressure, applied fields). Cells must sense and respond to fluctuations in these control parameters, and one outcome is the formation or dissolution of condensates. The distinctions between parameters are also not always clear: e.g., molecular affinities depend on physical fields such as salt. Nevertheless, viewing biology through the lens of condensation organizes diverse mechanisms into a tractable set of control parameters that can be used collectively to modulate the phase behavior of molecules inside, on, and outside of cells.

\subsection{Condensate Properties (“Responses”)}
Having considered the set of control parameters, we now discuss the measurable, quantitative responses, i.e., properties of the condensate that vary as these parameters are changed.

\begin{enumerate}
\item\textbf{Composition.} One such quantifiable response is the composition of a condensate, which comprises its proteins and nucleic acids, plus small molecules (ions, cofactors, metabolites, water). In cells, condensates are multicomponent and dynamic, typically containing dozens to hundreds of macromolecular species (proteins and RNAs, and in some cases DNA/chromatin). Unlike crystals, they lack fixed stoichiometry; constituent concentrations often vary with environment and time.
While stoichiometry can vary over time, the presence of an interface implies a composition that is persistently distinct from the surrounding environment. This non-fixed, exchanging nature---where molecules continuously partition in and out of condensates---complicates quantitative measurement of composition. Proximity-labeling proteomics and RNA profiling, along with curated component databases, have provided useful insights; though further advances are needed to obtain quantitative readouts of condensate-specific composition at high spatiotemporal resolution. Systematically quantifying how molecular composition shifts with control parameters (e.g., concentration, affinities, physical fields) will clarify what determines the makeup of condensates.

\item\textbf{Location.} A second key response pertains to where condensates form in cells. Mapping location was transformative—e.g., P granules in C. elegans embryos, whose posterior enrichment arises from spatially regulated dissolution/condensation (Box~\ref{box}), linked phase separation to development. Site-specific formation also encodes mechanism and function: condensates on microtubules often reflect scaffolded nucleation or active transport; membrane-proximal condensates (e.g., at junctional domains on cell membranes) can mediate signaling and imply specific protein–lipid interactions; nuclear-restricted condensates point to local concentration thresholds, binding landscapes, or ionic screening distinct from the cytoplasm. Systematically quantifying the spatial distribution of condensates, therefore, serves as a constraint in our deciphering of their mechanism and function.

\item\textbf{Timing (kinetics).} Temporal dynamics of condensates, i.e., timing or kinetics, is another measurable response. This concerns how fast condensates nucleate, grow, exchange components, age (waiting-time–dependent hardening), and dissolve, as well as when these processes occur. Because condensation is collective, multiple coupled timescales coexist: molecular binding or unbinding; chemical conversion; molecular diffusion; viscoelastic network relaxation; nucleation-and-growth versus spinodal demixing; growth or coarsening (i.e., coalescence or Ostwald ripening); and possible kinetic arrest or gelation. Disentangling them is difficult, but mapping them is diagnostic, revealing mechanisms: which processes occur, whether active remodeling (e.g., by ATPases, chaperones, helicases) maintains the state, and if coarsening is arrested by gelation. Timing also includes when condensates appear within physiological cycles (cell cycle, development, circadian time, aging) and how control parameters (e.g., physical stimuli) reshape lifetimes and dissolution rates. By mapping condensate kinetics alongside the timescales of other cellular processes, we can identify plausible links and generate testable hypotheses. Quantifying condensate timescales, therefore, enables predictive, mechanistic models.

\item\textbf{Material properties (mechanics).} Another key condensate response is material properties—the mechanical behavior of condensates (viscoelastic moduli, viscosity, surface or interfacial tension, gelation and yield behavior)—which quantify how condensates deform and flow under applied stress or strain. Although sequence-encoded interactions have long been linked to phase behavior, quantitative maps from sequence and environment to mechanical properties are only now emerging, enabled by improved rheology (particle-tracking or active microrheology, optical tweezers, droplet fusion and shape recovery, micropipette aspiration) in vitro and, increasingly, in cells. These parameters shape aging or maturation, control fusion or coarsening and fission, and determine how condensates respond to mechanical cues and, in turn, remodel their surroundings. Interfacial phenomena can even promote surface-mediated solidification, with disease implications. Systematically linking control knobs to material properties will be invaluable for both physiology and engineering.

\item\textbf{Internal organization.} Condensate internal organization also varies with control parameters. Here, we consider both the connectivity of molecules---often termed microstructure or substructure, which emerges on molecular length scales---and the formation of emergent condensate architectures, such as multiphasic organization, which occurs on larger length scales. Molecular connectivity governs key properties, including molecular diffusion, local interactions (e.g., enzyme kinetics), and information transfer within condensates (e.g., mechanical stress propagation). For instance, interfacial phenomena, including aging, have motivated comparisons of molecular organization at condensate surfaces versus cores. At larger length scales, interfacial effects can give rise to multiphasic condensates comprising multiple coexisting liquid phases. A well-studied example is the nucleolus, where distinct subphases within the condensate correlate with function and reflect underlying physical factors such as interfacial free-energy differences. Ultimately, internal organization shapes condensate mechanics and dynamics and provides insight into both function and the physical processes that regulate it.

\item\textbf{Morphology.} Lastly, control parameters also drive differences in condensate size and shape. Early work emphasized spherical droplets, but morphologies span 2D wetting layers on membranes and anisotropic or multiphase bodies. Crucially, shape only alludes to material state; spherical morphology does not guarantee liquidity. Condensates can be dynamically arrested or gel-like yet remain round, either because interfacial tension dominates or because a spherical liquid droplet has aged into a solid-like state while retaining its original shape. Small condensates also often appear spherical due to optical resolution limits. Quantitative readouts are therefore needed to infer mechanics. Size is likewise informative but biased: conventional imaging detects larger condensates and can miss nanoscale clusters that precede coarsening (e.g., in stress-granule assembly \cite{Ge2025}). Size and shape distributions, aspect ratios, contact angles, and their evolution over time reveal formation pathways (nucleation-and-growth vs spinodal; coalescence vs Ostwald ripening) and active remodeling.
\end{enumerate}

So far, we discussed the control parameters and the condensate responses they elicit at the single-condensate level (\figref{fig:concept}). One can also consider collective responses across many condensates. In some systems, condensate number is functionally critical (e.g., centrosomes), whereas in others numbers may be less informative (e.g., nucleoli). Furthermore, condensate count is not an independent response; it emerges from nucleation, growth, coalescence/fission, and dissolution, and should be interpreted alongside size and spatial distributions. It is also important to stress that many different sets of control parameters can result in similar condensate responses. Additionally, since control parameters vary in space and time, the resulting response space is inherently high-dimensional.

\section{What condensation can achieve in, on, and outside of cells}
\label{sec:biology}

Physical principles explain that molecules with sufficiently strong, multivalent interactions will separate into coexisting phases. This occurs with folded proteins (e.g., eye-lens $\gamma$-crystallins, lysozyme) as well as with highly disordered conformations (e.g., intrinsically disordered regions) inside cells, on the plasma membrane, and outside of cells. Various parameters of this process, such as timing, location, component ratios, internal structures, and material properties can be regulated by cellular mechanisms (\figref{fig:concept}). In turn, the emergent properties of resulting condensates greatly influence the cellular space, which can be utilized by cells or researchers for functional benefit. In this section, we discuss (i) how condensation impacts molecules and cells; (ii) how evolution may have harnessed condensation to control cellular processes; and (iii) how researchers can leverage condensation.

\subsection{Effects of condensation on the cellular space}

Condensates profoundly affect the cellular organization in at least three different ways:

\begin{enumerate}
\item\textbf{Composition and stoichiometry.}
Condensates create a unique physicochemical environment that causes certain molecules (including macromolecules, ions, and small molecules) to partition into, be excluded from, or accumulate at the condensate-dilute phase interface to varying extents. This process then modifies the composition or stoichiometry of molecules outside the condensates, potentially generating a physicochemical or compositional gradient across the cellular space.

\item\textbf{Conformation, interaction, and dynamics of molecules.}
The condensate composition (point 1) creates unique solvent properties within condensates, at their interface, and in their surrounding environment. This influences the diffusional and conformational dynamics of molecules—such as how they move, 'breathe,’ or fold—both inside and at the interface of the condensates, which in turn influences their intra- and inter-molecular interactions. For instance, proteins that adopt relatively compact conformations in dilute solution due to strong intramolecular interactions can exchange these interactions for more favorable intermolecular interactions within the condensate, causing them to adopt more open conformations individually. The surface of condensates also provides a distinct interface that impacts the dynamics and conformation of molecules. At this interface, proteins can form different intermolecular arrangements; emerging evidence suggests that the condensate interface can promote the formation of amyloid-like inter-protein beta-sheet transitions and condensate hardening.  

\item\textbf{Physical effects.}
Depending on their composition, size, location, and material properties, condensates can obstruct, exert forces on, or ‘glue' nearby structures such as molecules, membrane-bound structures, the cytoskeleton, chromatin, and other condensates. By occupying space as selective compartments, condensates can also function as crowders with varying degrees of physicochemical and material properties, which can alter local viscosity and slow or restrict the diffusion of macromolecules in their vicinity.
\end{enumerate}

These physically intrinsic and thus ‘unavoidable’ effects of condensates may influence cellular processes, or the processes must be resilient against them. We note that what we discussed in this section is not an exhaustive list of what condensation can achieve in biology. Additionally, the examples listed here are not mutually exclusive but closely interconnected, influencing one another and being influenced by other processes (i.e., the control ‘knobs’ mentioned in the previous section and \figref{fig:concept}).

\subsection{Ways condensation may be used to control cellular processes}

How might cells leverage these effects for their functional benefit? Below, we list examples of roles that condensates might play in cellular processes, based on principles of biomolecular interactions and condensate properties, along with experimental evidence from biology (Box~\ref{box}). 

\begin{enumerate}
\item\textbf{Spatial organization.} Condensates concentrate specific molecules in particular locations. The formation of a condensate thus allows cells to regulate their structure and provide spatiotemporal cues for downstream processes. For example, condensate microenvironments can organize various cellular structures—from cytoskeletal networks (such as the actin cortex~\cite{Yan2022a} and mitotic spindle~\cite{Oriola2020}) to membrane-associated structures (like membrane contact sites~\cite{Kim2024b} and signaling clusters that regulate immune responses~\cite{Case2019a})—in highly tuned hierarchies that lead to different morphologies. Additionally, condensation can control where signaling occurs (e.g., postsynaptic densities~\cite{Chen2020} and clustering of polarity factors~\cite{lasker2020}) or how cells are connected (such as tight junctions~\cite{Pombo-Garcia2024}), thereby influencing communication between cells and in turn, the organization of cells and tissues.

\item\textbf{Tuning biochemical reactions and pathways.} Condensation can provide spatiotemporal control over biochemical reactions, either to facilitate or inhibit them, as well as influence stochastic processes like nucleation (e.g., actin polymerization~\cite{Yan2022a}). Beyond individual reactions, this control can also tune pathway selection (e.g., branched pathway ‘channelling’ in metabolism~\cite{MartinezCalvo2024}) and separate interfering or competing processes. This can be achieved by concentrating or depleting specific molecules at certain locations. Additionally, the local chemistry within condensates can contribute, e.g., by modulating conformations and interactions. 

\item\textbf{Buffering.} Beyond a critical concentration, the ‘extra’ molecules can partition into condensates, leaving the effective concentration outside unchanged. Therefore, condensation may help buffer chemical processes and regulate gradients and fluctuations in molecular concentrations. However, the partition coefficient in biomolecular condensates is not always constant, %
and predicting the saturation concentration within the cellular milieu is not straightforward. %

\item\textbf{Force transmission and mechanics.} Condensation can generate and transmit forces that displace chromatin or drive chromosome segregation. Condensates can also serve as glues or plugs to support repair, as seen in DNA end synapsis during double-strand break repair~\cite{Chappidi2024}, and in the repair of ruptured lysosomal membranes~\cite{bussi_stress_2023}. The condensate solvent environment may also influence the structure or arrangement of polymers, e.g., by promoting microtubule aster formation over bundling and vice versa~\cite{MontenegroGouveia2018}.

\item\textbf{Information processing and transduction.} Condensation provides a highly sensitive spatiotemporal response, transferring molecular information (e.g., hydrophobicity, aromaticity, charge, and sequence pattern) to the cellular level across sharp decision boundaries. Therefore, condensates can be used to detect and respond rapidly to changing environmental conditions, such as temperature, light, or chemical signals (e.g., hormones) during plant or animal development and stress responses (e.g., starvation, dormancy, latency) as observed in bacteria, seeds, tardigrades, and marine animals. Condensation can also transfer information (e.g., domains of epigenetic modifications~\cite{mukherjee2024selforganisedliquidreactioncontainer}) or increase variability (e.g., random placement of crossovers during meiosis~\cite{morgan_diffusion-mediated_2021,zhang_crossover_2025}). These responses, processed or translated by condensation, are nonlinear, similar to aggregation processes but different in that they are reversible.
\end{enumerate}

Do cells leverage condensation to enhance their fitness and survival? Condensates are often associated with specific processes based on their composition or localization. It is tempting to speculate that they directly affect these processes. If so, certain ‘features’ of condensates might be conserved or reimagined through evolution to serve similar functions. These questions and possibilities are still actively studied and lack definitive answers. The answers will likely vary by case rather than apply universally. It is also possible that condensation has little to no impact on the process it is associated with or on the fitness of cells or organisms, especially in specific contexts. Regardless of the case, the presence of condensates will influence the cellular environment, which may need to be managed by the cells. It is possible that cellular mechanisms even suppress condensation to avoid such effects.

\subsection{How researchers can apply what we learn about condensates}
Understanding the properties and responses of condensates, along with their effects on cellular space and potential functions, offers scientists exciting opportunities to apply these principles to study cellular processes or to design a new generation of chemical factories or therapeutics. 

\begin{enumerate}
\item\textbf{Probe biological functions.} Using principles of condensation, we can develop methods to control different aspects of condensates, helping us better understand their roles in biological processes and diseases. Specifically, it will be useful to modulate condensation (formation or dissolution of certain molecules), condensate composition (concentration or depletion), or their properties independently, without deleting or altering both copies of endogenous proteins to separate the effects of modifying a condensate component from condensation. This may be accomplished through mutations, engineered peptides, or tags individually or in combination.

\item\textbf{Therapeutics.} As we identify condensates involved in disease states, therapeutics can be designed to target them, e.g., to dissolve them (e.g., to dampen oncogene transcription) or to modify their material properties (e.g., to restore healthy dynamics). 

\item\textbf{Designer condensates.} Synthetic condensates can be engineered to have specific properties that control targeted outputs, such as sequestering harmful components, turning signaling or reactions on or off, or generating new materials both inside and outside of cells, to probe biological functions or serve as therapeutics.
\end{enumerate}

When studying the potential roles of condensates in biology and in disease, it is important to recognize that condensation might not always promote or enhance a process; it can also suppress or inhibit it, even to the point of being toxic to cells. Therefore, it will be crucial to develop ways not only to induce condensation but also to prevent it. 

In summary, we suggest considering condensation as a physical tool accessible to both cells and researchers. Does it help to study your process of interest by evaluating and applying the properties of condensates discussed here? The concept of phase coexistence can be a powerful tool for predicting specific behaviors of biomolecules or for explaining biological phenomena through physics. For instance, new sights may arise from considering partitioning rather than stoichiometric binding or unbinding, or examining phase boundaries and solvent environments rather than physical obstructions caused by occupancy. Emerging evidence indicates that this approach can be very effective (Box~\ref{box}). However, predictions and behaviors vary widely across systems, and there is no universal set of tests to ‘prove’ condensation; close collaboration among biologists, physicists, chemists, mathematicians, and engineers is thus essential.

\section{Challenges}
\label{sec:challenges}

While the biomolecular condensates research has made significant progress in advancing the biophysical understanding of condensates and their roles in cellular processes (Box~\ref{box}), it faces several challenges going forward. Some of these are laid out below under the headings of conceptual, methodological, and social challenges.

\subsection{Conceptual challenges}
\paragraph{Sensitivity to conditions and perturbations.} A central conceptual challenge has to do with the energy scale associated with condensation. By the very nature of condensates as structures that can form dynamically according to cellular needs, condensation must occur on the border of stability. Thus, energy scales for condensation are modest, typically on the order of a few $k_\mathrm{B}T$ per molecule, to allow for regulation via changes in conditions such as pH or temperature or via phosphorylation, etc. This, in turn, means that condensates are highly sensitive to both external conditions and to changes of condensate components, such as modifications or mutations. Many of the known challenges to understanding condensate biophysics and biology arise directly from this sensitivity:

\begin{enumerate}
\item The observation that condensates often behave differently in vitro from in vivo reflects sensitivity to the complex environment of the cell, including ions, and other small molecules, crowding by macromolecules, changing composition (see point b), the presence of active, energy-consuming processes, and more generally the fact that cells are not in equilibrium.
\item The subtle dependence of condensate properties on polymer sequence likely also reflects this sensitivity. Notably, evolution has fine-tuned the relevant protein and RNA sequences to condense and function in the exact conditions of the cell. Moreover, sequence features interplay with the conformational degrees of freedom of the unstructured regions in condensate components. As such, the internal organization of condensates can be quite complex and heterogeneous. Indeed, in view of the selective pressure of evolution on sequence details, nothing about natural condensates can be assumed to reflect “typical” homopolymer behavior. 
\item The small energy scales associated with condensation present a challenge to theory: the inevitable small inaccuracies in modeling can drastically change predictions. As such, theory remains very limited in its ability to predict condensate phase diagrams, physical and chemical properties, and functional behavior.
\item In cells, condensates can participate in positive or negative feedback loops, which act back on specific components, making their behavior additionally complex. For example, enrichment of kinases that phosphorylate certain components can modify condensate properties or even lead to condensate dissolution.
\end{enumerate}

\paragraph{Multicomponent nature of condensates.} Another conceptual challenge, of particular relevance to condensates in cells, is that the exact components of condensates are often not known, and even the known components may be chemically modified in unknown ways. In addition to the dominant phase-separating components, often called “scaffolds”, in vivo condensates may contain hundreds of low abundance components, also known as “clients”. Rather than being passive hitchhikers, clients can affect the stability and dynamics of condensates, again reflecting the overall small energy scales involved in condensation.

\paragraph{Linking condensate properties to function and dysfunction.} The field of condensate research is presented with both opportunities and challenges in translating fundamental biophysical insights into clinically relevant health applications. A significant hurdle lies in establishing clear causal relationships between condensate dysfunction and disease pathogenesis, as many condensate-related disorders may involve subtle alterations in phase behavior that are difficult to detect and quantify in living systems. Additionally, developing therapeutic strategies that can precisely modulate condensate properties without disrupting essential cellular functions remains challenging. The complexity of condensate composition noted above—often involving hundreds of different proteins and nucleic acids with varying interaction strengths—makes it challenging to predict how genetic mutations or environmental factors will affect condensate behavior and subsequently contribute to pathological conditions such as neurodegenerative diseases or cancer. Of particular relevance to condensate-mediated disease, many condensates are observed to age, i.e., to become more viscous and solid-like with time. It remains a significant challenge to predict which condensates are subject to aging and on what time scales. While beta-sheet formation is implicated in some aging processes \cite{Emmanouilidis2024}, other mechanisms may also be involved, e.g., gradual rearrangements into more favorable bonding configurations. As such, aging or its lack thereof may depend on the recruitment into condensates with interacting molecules that help prevent beta-sheet stacking or other stable arrangements such as chaperones or post-translational modification enzymes. Insofar as aging may depend on the nucleation of rare structures, rates of aging are likely to be extremely sensitive to heterogeneity and fluctuations within condensates.

\paragraph{De novo functional condensates.} Finally, a major conceptual challenge is the design of novel functional condensates. Since condensates are ubiquitous in nature, it seems evident that creating condensates de novo will be a powerful tool for cellular engineering. Particular applications include redirecting metabolism toward desired products (biofuels, fine chemicals) via co-clustering of enzymes within a desired pathway, and condensates that act as biosensors to report on cellular states. However, just getting desired molecules to condense inside a cell is not enough to guarantee proper function. A worthy goal will be the design of condensates with the right components (both who's in, and who's out), size, localization, dynamics (forming when and where needed), physical properties (e.g., exchange rates, viscosity, surface tension), and possibly interactions with other organelles and cellular structures.

\subsection{Methodological challenges}
\paragraph{Probing microstructure of condensates.} One of the main technical challenges in the study of condensates is the difficulty of probing structures of $\sim\SI{100}{\nano\meter}$ scale. Atomic-scale structures can be probed by X-ray crystallography, NMR, and electron microscopy, while macroscopic structures can be probed by light microscopy. Natural condensates often occupy the $\sim$hundreds of nm “no man’s land” in between. Moreover, because condensates are often amorphous, tools such as cryo-electron microscopy and tomography typically only yield very rough pictures of condensates with little information about detailed structure (albeit with some exceptions \cite{he_structural_2020,Rog2017}). To learn more about internal condensate structure, alternative approaches include single-particle tracking and reporters of local organizations such as FRET, FLIM, and EPR. Since many probes of condensates require labeling, e.g., with fluorescent proteins or other fluorophores, it is important to keep in mind that these tags can change phase behavior in unpredictable ways. 

\paragraph{Composition and molecular states in condensates.} Another technical challenge, touched on above, is the difficulty of knowing precisely what is inside a condensate. This is particularly true in cells, where the primary scaffolds may exist in a variety of modification states, and there are likely to be multiple clients and miscellaneous interlopers. Biochemical isolation of condensates faces the challenge of preventing their full or partial dissolution, e.g., during standard fractionation procedures. Nonetheless, it is notable that condensate isolation followed by mass spectrometry analysis has yielded insights in some cases, as have proximity-labeling approaches \cite{youn_high-density_2018, markmiller_context-dependent_2018}. For in vitro studies, it is still challenging to know the exact densities and ratios of condensate components, or which molecules interact with which others and how; even for single-component condensates, the range of polymer conformations is generally unknown and may vary, e.g., between the surface and bulk, both in vivo and in vitro.

\paragraph{Condensate physical properties.} Measuring the physical properties of condensates, like surface tension, viscosity, or electrical charge, presents its own challenges. Direct measurement of these properties in vivo is generally out of reach. FRAP has been used productively both in cells and in vitro to measure exchange rates of condensate material and diffusion coefficients within condensates \cite{Taylor2019}, although one must keep its limitations in mind as well \cite{mcswiggen_evaluating_2019,taylor_quantifying_2019}. Chemical treatments that dissolve condensates (like 1,6-hexanediol) are often non-specific and may have off-target effects. Genetic manipulations can change in vivo condensate properties, but these take time to manifest and may trigger compensatory responses. Optogenetic techniques hold promise but will require additional development to yield quantitative measures of in vivo condensate properties. In vitro studies allow more direct probing of condensate properties but are typically limited to a small number of components and do not necessarily reflect in vivo composition or conditions. (A perhaps obvious point is that many proteins or RNAs will phase separate in vitro at high enough concentrations, but that is not enough to establish meaningful phase separation in vivo.) Growing interest in reproducing condensate dynamics – such as nucleation, coarsening, size control, dissolution, etc. – in vitro also faces the need for high spatial and temporal resolution measurements and additional components, such as energy sources, when active processes are at play. The latter include post-translational modifications that may regulate condensate molecular rearrangements or dissolution in vivo and are difficult to reproduce in vitro.

\paragraph{Sequence-resolved modeling of condensates.} As remarked above, condensates’ intrinsically dynamical character and their fine-tuning by evolution present substantial challenges to theory and modeling. As noted, the proteins and RNAs that form condensates cannot be expected to behave like homopolymers or even random heteropolymers, limiting the ability of generic polymer models, such as Flory-Huggins theory, to address quantitative questions. In some cases, sequence-dependent analytical theories such as random phase approximation and renormalized Kuhn length can quantitatively rationalize experimental data and even make high-throughput predictions for sequence-specific partitioning into condensates \cite{Wessen2025}, but these theories are still limited in their ability to capture structural and energetic details \cite{Lin2023}. While molecular and field-theoretic simulations have proven valuable in studying condensates, the wide range of relevant timescales presents significant difficulties. In particular, all-atom simulations, which capture the nuances of specific sequences of residues, are typically limited to time scales of microseconds for small condensates. Abstract coarse-grained models are useful to test conceptual points, but cannot be mapped one-to-one on specific proteins or RNAs. Coarse-graining at the residue level is an attractive compromise for modeling the condensation of specific sequences, but it only approximately captures detailed interactions and dynamics and does not yet reach the long time scales associated with aging.

\paragraph{Deciphering the biological function of condensates.} Finally, an elephant in the condensate room is the question of what exactly does condensation do for function? The challenge here is to separate condensation per se from other processes. Mutations or other perturbations that abrogate condensation may affect function in multiple ways: For example, if condensation follows from phosphorylation of a particular protein, a mutant that lacks phosphorylation sites won’t condense – but the loss of function might instead be due to the lack of phosphorylation. Moreover, assays for biological function are often quite indirect, occurring downstream of condensate-mediated activities, adding uncertainty to inferring the role of condensation. A related challenge is that the function of condensation in cells may itself be indirect. For example, condensation could be employed to localize factors within a cell, or to keep out unwanted factors from a particular location. These are important functions, but have little to do with the direct ability of condensates to create high densities of particular factors. In sum, to address the function of condensation, a challenge for the field will be to develop perturbations that affect condensation while minimally perturbing other properties of the molecules involved.

\subsection{Social challenges}

\paragraph{Condensate studies straddle physics and biology.} A major “social” challenge to the study of condensates is the need to communicate and collaborate across disciplinary barriers. The research area draws from cell biology, biophysics, biochemistry, structural biology, polymer physics, and materials science, each with distinct experimental approaches, theoretical frameworks, and preferred conferences and journals. For example, biophysicists may focus on phase diagrams and scaling relations while cell biologists emphasize mutant studies and functional outcomes. Such differences make it difficult to establish close collaborations with shared research questions. There are many illustrative examples in biology of successes of integrating experiment and theory, e.g., going back to \citet{luria_mutations_1943}, but it is still the exception for experimental labs to involve theory prior to doing experiments. This is equally the case in studying condensates, where theorists are typically faced with analyzing pre-existing data. A particular challenge will be to convince experimentalists to involve theorists at earlier stages, e.g., in choice of model system and experimental approach, and with an eye toward addressing conceptual questions and interpreting the underlying biophysics. On the flip side, it will be important for theorists to engage with the details of the experiment and the underlying biology of specific systems. There are only a limited number of universal conceptual questions in biomolecular condensates – likely not enough to keep all the interested theorists occupied – but there is a wealth of questions regarding how specific condensates exploit physics to robustly perform their unique functions. Thus the future of biophysical theory in the study of condensates is likely to be increasingly “biological”. To this end, bringing together experimentalists and theorists in conferences or workshops with specific opportunities to develop collaborations will be productive. Since ease of communication is one key to making such collaborations happen, it should be a priority to educate members of each group in the language of the opposite side.

\paragraph{Less hype, more rigor.} While the concept of condensates is successfully rewriting cell biology textbooks, there is some danger of overhype and backlash. The excitement around condensates has led to claims that phase separation explains many previously mysterious cellular phenomena. The pressure to publish or rapidly commercialize is particularly high in cases of potential therapeutic relevance of condensation, e.g., in neurodegeneration and cancer. However, this enthusiasm around condensates may sometimes outpace rigorous experimental validation, potentially creating skepticism. It is thus important for high standards to be maintained in the face of all these pressures.

\paragraph{Future job and funding opportunities.} For the field to flourish, it is also essential that researchers focused on condensates be able to find jobs and get funded. Universities don’t have condensate departments, so academic researchers need to find jobs in departments such as Molecular Biology, Biochemistry, Chemistry, or Biophysics. Will the more biological departments be willing to hire scientists interested in biophysical or theoretical aspects of condensates? Will funding agencies be willing to support condensate research that cuts across traditional disciplines and model systems? Will biotech and pharma companies step up to pursue condensate research? Only time will tell...

\section{Outlook}
\label{sec:outlook}

\begin{figure*}
    \centering
    \includegraphics[width=0.8\linewidth]{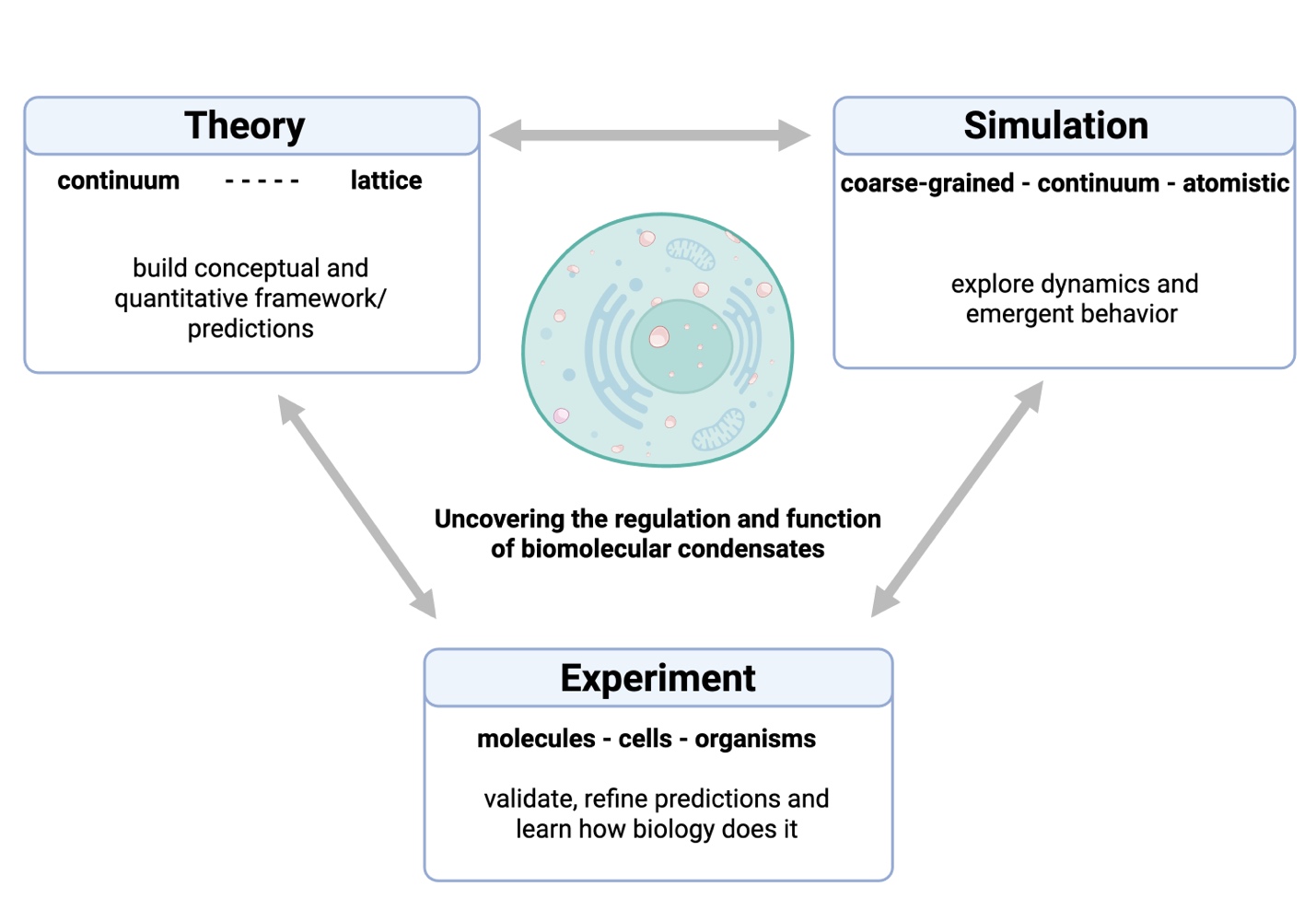}
    \caption{
    Understanding biomolecular condensates and their biological consequences will require integrating theoretical, numerical, and experimental approaches. Each approach offers extensive tools, such as chemically non-specific, composition-dependent, and sequence-specific theories, as well as exploring similarities and variability across different size and time scales (e.g., evolution) in experiments. Figure generated using BioRender.
    }
    \label{fig:disciplines}
\end{figure*}

Biomolecular condensation has emerged as a fundamental principle of cellular organization, transforming how we think about the spatial and temporal organization of molecules inside cells. The preceding sections have laid out how condensates can be understood in terms of control parameters, measurable responses, and their diverse roles in cell physiology, as well as the challenges faced in this endeavor. The field started from simple premises and has matured to provide deep new insights into cell biology. However, to embrace the full biological complexity, we must tackle conceptual, technical, and social challenges that span scales and disciplines.

\paragraph{From phase separation to complex supramolecular organization.} Phase separation provides a powerful framework to understand the segregation of molecules in distinct phases separated by an interface. Cellular organization is typically more complex. In the out-of-equilibrium environment of cells, structures and phenomena emerge that go well beyond segregated phases. Phase separation is a generic physical concept, like diffusion, fluid dynamics, and mechanics, that is key to identifying the principles of cellular organization. However, many challenges lie ahead in obtaining a full understanding of the physics of structures such as the nuclear envelope, the spindle, and chromatin organization.

\paragraph{Dynamics and control in living cells.} Condensates are active, non-equilibrium structures shaped by chemical reactions, mechanical forces, and energy fluxes. Future work will focus on how cells regulate nucleation, growth, coarsening, and dissolution of condensates on physiological time scales, and how feedback from metabolic activity, cytoskeletal organization, or signaling pathways tunes these dynamics. A particular goal is to understand how robustness and adaptability could emerge from the interplay of active processes with phase separation.

\paragraph{Linking physical organization to biological function.} Condensates influence gene regulation, signaling, metabolism, and mechanical properties of cells and tissues. Determining when condensation is causal for these functions—rather than merely correlated—remains a major frontier. New perturbative approaches that can specifically modulate condensate formation or material state, combined with quantitative readouts of biochemical and mechanical outcomes, will be key to establishing causal links.

\paragraph{Condensates in development, evolution, and disease.} The role of condensates across scales—from subcellular patterning to tissue morphogenesis and organismal development—presents rich opportunities. Exploring how condensate properties are selected and maintained by evolution, and how their dysregulation could contribute to aging, neurodegeneration, and cancer, will deepen both basic and biomedical insights. Identifying conserved physical design principles may also clarify how different organisms exploit condensation for robustness and adaptability.

\paragraph{Engineering and medical opportunities.} A deeper physical understanding of condensate formation in the multicomponent environment of the cell could enable the rational design of synthetic condensates with specified composition, dynamics, and material properties. Such designer condensates could rewire metabolism, organize artificial organelles, or serve as drug-delivery systems. Conversely, precise interventions to dissolve or remodel pathological condensates offer new therapeutic strategies.

\paragraph{Convergence of disciplines.} Progress will depend on close interaction between physics, chemistry, and biology. Developing multiscale theory, high-resolution and high-throughput experimental tools, and data-driven analysis methods will require sustained collaboration and a converging and shared language. The study of biomolecular condensates in cells and organisms could bring about a new research field as a convergence of existing disciplines.

By integrating theory, simulation, and experiment, and by embracing the complexity of living systems (\figref{fig:disciplines}), condensate research could play a key role in opening new avenues in cell biology and also bring new approaches to control and engineer the molecular organization of life.

\textit{Acknowledgments---} Many colleagues have contributed to stimulating discussions on these topics over the years. This work was supported in part by grant NSF PHY-2309135 and the Gordon and Betty Moore Foundation Grant No. 2919.02 to the Kavli Institute for Theoretical Physics (KITP).

\bibliography{literature.bib}

\end{document}